\documentclass[12pt, letter]{emulateapj}
 \usepackage{epsf}
 \usepackage{epsfig}

\newcommand{\etal}{et al.}

\newcommand{\ie}{{\it i.e.}}

\def\deg      {{\ifmmode^\circ\else$^\circ$\fi}} 

\slugcomment{ }

 \shorttitle{CTI and galaxy shape measurement}
 \shortauthors{Jason Rhodes et al.}

\begin{document}


 \title{The effects of charge transfer inefficiency (CTI) on galaxy shape measurements}




\author{Jason Rhodes\altaffilmark{1,2},
Alexie Leauthaud\altaffilmark{3,4},
Chris Stoughton\altaffilmark{5},
Richard Massey\altaffilmark{6},
Kyle Dawson\altaffilmark{7},
William Kolbe\altaffilmark{3},
Natalie Roe\altaffilmark{3}}

\email{jason.d.rhodes@jpl.nasa.gov}

\altaffiltext{1}{Jet Propulsion Laboratory, California Institute of Technology, Pasadena, CA 91109}

\altaffiltext{2}{California Institute of Technology,  Pasadena, CA 91125, USA}

\altaffiltext{3}{Lawrence Berkeley National Laboratory, 1 Cyclotron Rd.
  Berkeley, CA 94720}

\altaffiltext{4}{Berkeley Center for Cosmological Physics, University
  of California, Berkeley, CA 94720, USA}

\altaffiltext{5}{Fermi National Accelerator Laboratory, Batavia, Illinois, 60510}

\altaffiltext{6}{Royal Observatory Edinburgh, Blackford Hill, Edinburgh EH9 3HJ, U.K.}

\altaffiltext{7}{Department of Physics and Astronomy, University of Utah, Salt Lake
City, UT 84112}




\begin{abstract}

  We examine the effects of charge transfer inefficiency (CTI) during
  CCD readout on the demanding galaxy shape measurements required by
  studies of weak gravitational lensing.
  We simulate a CCD readout with CTI such as that caused by charged
  particle radiation damage in space-based detectors. We verify our
  simulations on real data from fully-depleted p-channel CCDs
 that have
  been deliberately irradiated in a laboratory.
  We show that only charge traps with time
  constants of the same order as the time between row transfers during readout
  affect galaxy shape measurements.  We simulate deep
  astronomical images and the process of CCD readout, characterizing the effects of CTI
  on various galaxy populations. Our code and methods are general and
  can be applied to any CCDs, once the density and characteristic release times of their charge trap
  species are known.
  We  baseline our study around  p-channel CCDs that
 have been shown to have   charge transfer efficiency  up to an order of
  magnitude better than several models of n-channel CCDs designed for space applications.
  We predict that for galaxies furthest from the readout registers, bias in the measurement of galaxy shapes, $\Delta e$,
  will increase at a rate of $(2.65\pm 0.02)\times10^{-4}\textrm{yr}^{-1}$  at L2 for accumulated radiation exposure averaged over the solar cycle.
  If uncorrected, this will consume the entire
  shape measurement error budget of a  dark energy mission surveying the entire extragalactic sky within about
  4 years of accumulated radiation damage.  However, software mitigation
  techniques demonstrated elsewhere can reduce this by a factor of $\sim10$,
  bringing the effect well below mission
  requirements. This conclusion is valid only for the  p-channel CCDs we have modeled;
   CCDs with higher CTI will fare worse and may not meet the
  requirements of future dark energy missions.
  We also discuss additional ways in
  which hardware could be designed to further minimize the impact
  of CTI.

\end{abstract}



\keywords{cosmology: observations -- gravitational lensing -- large-scale
structure of Universe}



\section{Introduction}

The past decade has seen enormous changes in the field of cosmology.
A concordance cosmology in which the expansion of the universe is
accelerating has been accepted (Spergel \etal\ 2007).  This
accelerated expansion was first demonstrated by observations of SN Ia
(Riess \etal\ 1998; Perlmutter \etal\ 1998) and has been confirmed
with other probes in recent years.  This startling aspect of our
Universe has prompted a wide variety of possible explanations (see,
e.g., Caldwell 2004) and considerable effort has gone into developing
concepts for dedicated space missions to probe the mysterious dark
energy thought to be causing this accelerated expansion.  These
missions, which include the NASA/DOE Joint Dark Energy
Mission\footnote[8]{See http://jdem.gsfc.nasa.gov} and ESA's Euclid
mission\footnote[9]{See http://sci.esa.int/euclid}, plan to use a variety of
probes in order to constrain the properties of dark energy.  It has
become widely accepted that weak gravitational lensing, the small
distortion of the observed shapes of background galaxies by foreground
dark matter, is one of the most powerful probes of dark energy,
provided that systematic effects can be controlled (Albrecht \etal\
2006). Space missions are attractive in large part due the greater
ability to control many systematic effects (Rhodes \etal\ 2004a).

The field of weak lensing by large-scale structure, or cosmic shear,
developed in parallel to the study of dark energy as the dominant
component of the Universe.  From the first detections of cosmic shear
a decade ago (Wittman \etal\ 2000; Bacon, Refregier \& Ellis, 2000;
Kaiser, Wilson, \& Luppino, 2000; Van Waerbeke \etal\ 2000), surveys
have grown in size and thus information content (see Hoekstra \&
Jain 2008 for a recent review).  The culmination of this effort will
be in the execution of the above-mentioned dedicated dark energy
missions, which plan to survey up to 20000 square degrees, the entire
extragalactic sky. It has become clear that control of systematic
effects, both observational and astrophysical, is of paramount
importance in making use of weak lensing as a probe of dark
energy. The subtle shape changes induced by weak lensing require
exquisite control of observational systematic effects, especially
knowledge of the telescope's point spread function (PSF). From the ground, thermal and gravity load-induced fluctuations in the telescope can change the PSF, and the atmospheric seeing both broadens the PSF and makes the PSF unstable on the timescale of astronomical exposures. These effects can be largely or completely mitigated by making observations in a thermally stable environment above the atmosphere.

The control of systematics is a large factor in the
drive towards a dedicated space mission.  However, the harsh radiation
environment of space has the potential to introduce an observational
systematic effect due to charge traps created in CCD detectors by
impacts from charged particles.  These defects trap charge
(either electrons in so-called n-channel CCDs or holes in p-channel CCDs) as
charge is clocked across pixels toward the readout registers.
 When
the charge is subsequently released from the trap, it shows up in a
neighboring pixel, thus creating a trail along the readout
direction.
These trails obviously change the observed shapes of the
galaxies in the images.  These shape changes are coherent across the
image, thus mimicking a weak lensing signal caused by dark matter. The
degradation of charge transfer efficiency (CTE, this quantity is one
minus the CTI, or charge transfer \emph{inefficiency}) due to these
radiation-damage induced traps has been observed in all Hubble Space
Telescope (HST) imaging cameras: the Wide Field Planetary Camera 2
(WFPC2; Dolphin 2009), the Space Telescope Imaging Spectrograph (STIS;
Goudfrooij \etal\ 2006), and the Advanced Camera For Surveys (ACS;
Sirianni \etal\ 2005) and has hampered the sensitive shape
measurements needed for weak lensing with those cameras (Rhodes \etal\
2004b; Schrabback \etal\ 2007; Rhodes \etal\ 2007).  CTE degradation is a particularly difficult effect to correct for because its non-linear nature means that high signal-to-noise (S/N) stars which are typically used in weak lensing for PSF modeling will be affected less than the low S/N galaxies whose shapes are being measured. Thus, typical PSF deconvolution techniques are complicated by the effects of CTI. Thus, it is clear
that future space weak lensing missions will need to minimize CTI due
to radiation damage and have CCDs that are sufficiently well
understood to allow for mitigation of the CTE degradation that does
occur.

In this paper we carry out a quantitative analysis of the effects of CTI in CCDs on the analysis of
galaxy shapes for weak lensing, and explore techniques to mitigate the shape distortions due to trailing charge.
We have developed a detailed model for the effect of charge traps based on data from irradiated CCDs,
and applied it to simulated galaxies.
We quantify the effects on measured galaxy shapes in simulated data as a function of galaxy size,
CTI, and S/N.  We base our analysis on data from p-channel CCDs fabricated
at Lawrence Berkeley National Laboratory (LBNL) and irradiated with protons at the
LBNL 88'' cyclotron (Dawson et al 2008; hereafter D08).
However, our methods are general and can be applied to any CCDs if the density and time constants of the charge traps are known.


This paper is organized as follows.  \S\ref{sec:cte} gives a brief
overview of how charge is transferred between pixels during CCD readout
and how different types of CCDs have performed after radiation damage.  \S\ref{sec:code} describes the code we have developed to mimic
the effects of CTI on CCD readout.  In \S\ref{sec:validation} we
validate that code by showing that it can reproduce the effects of CTE
degradation as measured on real, irradiated LBNL CCDs.  We apply our
code to simulated astronomical images and detail the effects of CTI on
the shapes of galaxies in \S\ref{sec:simulations}.  We examine how
this will effect the future space missions and give recommendations
for mitigating the effects in \S\ref{sec:future}.  Finally, we offer
concluding remarks in \S\ref{sec:conclusions}.


\section{Background}\label{sec:cte}

\subsection{CCD readout and charge transfer}

\begin{figure*}[tb]
\plotone{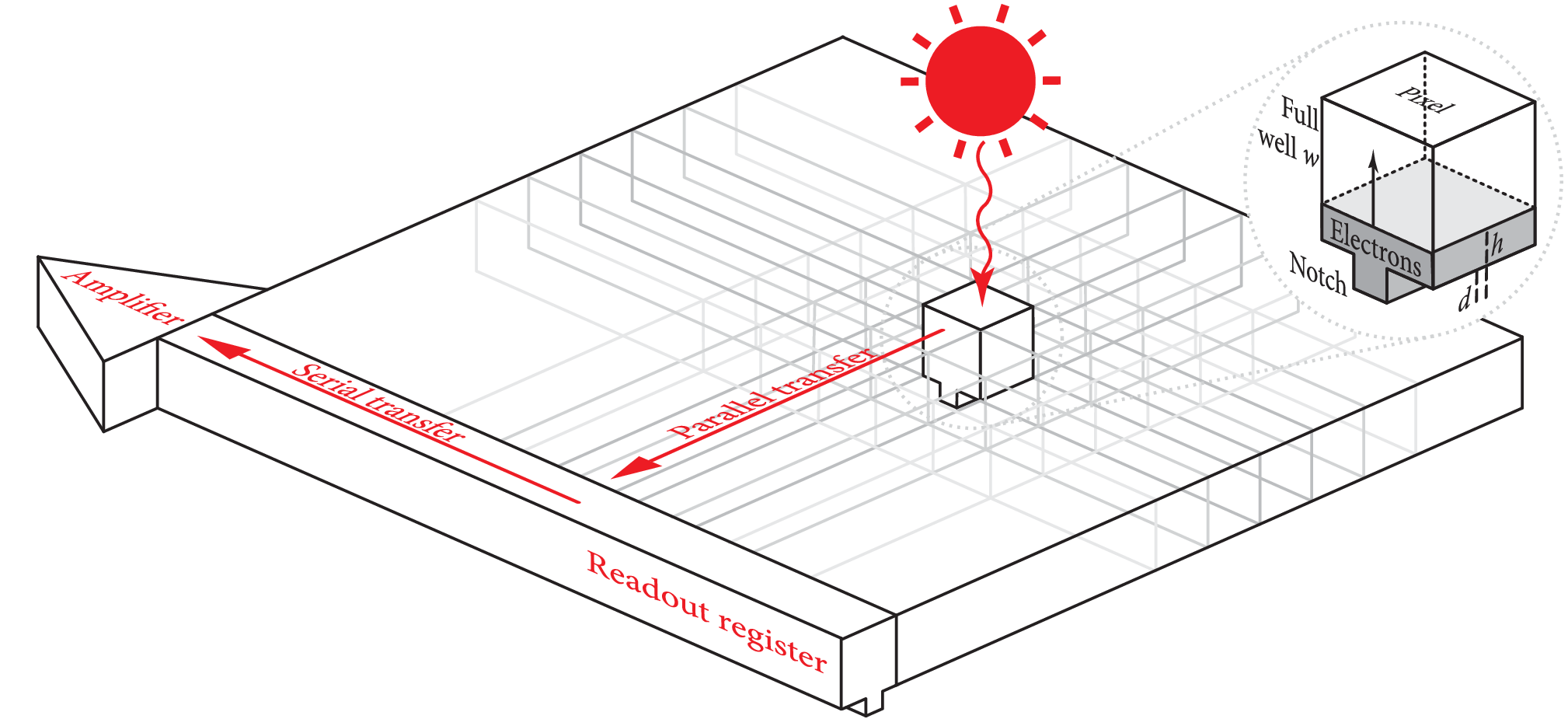}
\caption{Cartoon illustrating the CCD readout process.}
\label{fig:cte_diagram}
\end{figure*}

During exposure, a photon incident on a CCD generates an electron-hole pair.
In n-channel CCDs, the electrons drift into the potential well
of the nearest pixel, which is created
 by an electrostatic potential gradient within the substrate. In p-channel CCDs,
 holes are  collected
 instead, but we shall not distinguish between the two mechanisms hereafter.
 Charge (electrons or holes) accumulates in
a well-defined volume, outside which the density falls rapidly to zero (Hardy, Murowinski \& Deen 1998, Seabroke \etal\
2008). We are  most concerned with the cross-sectional area of this cloud, which expands as a monotonic function of
the amount of charge $n_e$. As illustrated in Figure~\ref{fig:cte_diagram}, we parameterize this as an effective
height $h$ within the pixel; the variable need not really be the height, but that is a useful one-dimensional function for the
purposes of explanation.

Some CCDs contain a supplementary buried channel or ``notch'' constructed at the bottom of the potential well. This is a
small region of slightly lower potential and, like the channel in an artificial river bed designed to improve the flow
of a small amount of water, it will concentrate the first few electrons (or holes). The notch reduces the number of traps the charge packet is exposed to,
by confining the charge to a smaller region.
We model the notch by setting the height $h$ to zero below some notch depth $d$, and
\begin{equation}
h(n_e)=\left( \frac{n_e-d}{w-d} \right) ^{\alpha}
\end{equation}
above the notch, where $w$ is the full well depth (the total amount of charge that will fit in a pixel before it
overflows) and $\alpha$, which depends on the construction of the potential but which is typically $\sim0.5$ (see for instance, Chiaberge \etal\ 2009 and Mutchler \& Sirianni, 2005).
Setting $h$ to zero below the notch is an approximation because it is possible that there are some traps in the notch.
At any rate, the CCDs we model in this paper have no notch  in the  imaging region and thus this approximation does not affect the results presented here.


At the end of the exposure, the charge must be moved to the amplifier electronics at the edge of the CCD to be read out
and digitized. Each row of charge
is first shuffled one pixel in the parallel direction, towards the serial
readout register. This
is typically accomplished using a 3-phase clock,
as is the case for the CCDs described here (Janesick 2001), but can be approximated as a single operation
(for a discussion of the consequences of this approximation, see Massey \etal\ 2009).
Charge from the bottom-most row
is transferred into the serial readout register
and then shuffled using the same technique but in a perpendicular direction.
The charge from each pixel is shifted onto a capacitor connected to an amplifier and
then `counted' by being sensed as a voltage and digitized.
Read noise is the shot noise on this voltage and can be reduced by lengthening the
 sampling time, and thus slowing the readout rate.
 This process is then repeated once for each additional row of pixels:
shuffling one row in the parallel direction and then through the serial readout register.


As the charge is transferred from pixel to pixel,
charge traps due to defects or impurities in the Si lattice temporarily capture passing charge, and
release  it
after some delay. The typical capture time is effectively instantaneous (and if it is not, the lower
capture rate can be equivalently modeled as a lower density of charge traps). The probability of release is governed by
an exponential decay, with a characteristic time constant
that depends upon the properties of the lattice and
impurities, as well as the operating temperature of the detector (Shockley \& Read 1952, Hall 1952). Several different
species of charge traps may be present in any given device, with different characteristic release times
 $\tau$,
and densities $\rho$.
 If the captured charge is released after its charge cloud has been shuffled along,
the released charge appears as a faint trail behind the main charge packet.

Trapping and trailing can occur in both the parallel and serial directions. Each column of pixels is independent and has
a unique set of charge traps, while all of the rows share the same traps along the serial register.  Note that even
though the pixels and transfer mechanisms are physically similar in the two directions, the time per pixel transfer is
typically $\sim 10^3$ shorter in the serial direction (D08). We demonstrate in \S\ref{sec:validation} that only
charge traps with release times roughly similar to the clocking (time between transfers) affect galaxy shapes. Thus, separate
species of charge traps can be important for parallel and serial transfers.

\subsection{CTE Effects in Irradiated CCDs}

Thick, fully depleted p-channel CCDs  have several advantages over conventional
thin, n-channel CCDs, including enhanced quantum efficiency at near-infrared wavelengths, reduced
fringing at near-infrared wavelengths, and significantly less
degradation of CTE with a given accumulated radiation exposure. This last effect is
particulary important for the measurement of subtle weak lensing-induced galaxy shape distortions. It arises due to the fact that the divacancy
traps that are primarily responsible for CTI in p-channel devices are more difficult to form than the phosphorous vacancy traps
that occur due to radiation damage in n-channel CCDs (Bebek \etal\ 2002; Janesick \& Elliott
1992; D08; Spratt \etal\ 2005).
We focus our analysis on p-channel CCDs precisely because it is more difficult to form traps in these types of
CCDs than in existing n-channel CCDs.

Marshall \etal\ (2004) compared the CTE responses of irradiated
p-channel LBNL CCDs  with n-channel CCDs.  These n-channel devices are designed for space applications and are the ones
used for the recently installed Wide Field Camera 3 on the HST.  They found that a notch implant in the channels
improved the CTE performance by a factor of 2 for both p-channel
and n-channel devices.  More importantly, they found that the CTE performance of p-channel devices is about an
order of magnitude better than that of n-channel devices after irradiation.
A re-analysis of the Marshall \etal\ data (Lumb 2009) indicates that the p-channel devices may only have about a factor of 3-8 better CTE (depending on the signal level, with the p-channel advantage being greater at low signal levels such as those expected in the images of faint galaxies). Likewise, a comparison  by Gow \etal (2009) found a factor of 7 improvement in tolerance for parallel CTI (and 15 for serial CTI) in otherwise similar p and n-channel devices.
Thus,  thick, fully depleted p-channel CCDs
are particularly attractive for a weak lensing space mission and we
use them as the baseline for this study.  We do, however, note that p-channel devices do not have the rich heritage that n-channel devices do, particularly in space applications.

LBNL has
developed radiation-hardened CCDs with the specific application of
dark energy missions in mind (Holland \etal\
2006). These CCDS are composed of $3512\times3512$ $10.5\mu$m pixels with 4 readout registers and were baselined for
the SuperNova Acceleration
Probe (SNAP), a JDEM concept (Bebek 2007). The SNAP mission, like other candidate
dark energy missions, would be at the L2 Earth-Sun Lagrange point, and
we use the radiation flux there as the baseline for the flux
experienced by a dark energy mission.  A dark energy observatory will
experience significant radiation exposure at L2,
primarily from solar protons.  Exposure to energetic protons leads to
degraded CCD performance due to bulk damage from non-ionizing energy
loss (NIEL) and charging of oxide layers from ionizing radiation.  Bulk
damage in the Si lattice is the dominant effect that manifests itself
through increased CTI, increased dark current, and isolated hot
pixels.  Of these effects, an increase in CTI is the most likely to
introduce systematic errors to a weak lensing survey.


\section{Software Algorithm to Mimic CTI}\label{sec:code}

Our model for CTI is that inefficient charge transfers are caused by
discrete charge traps embedded in pixels.  These charge traps can
capture and release single electrons (or holes). Each pixel of the detector contains a number of charge traps,
that have a mean density $\rho$. Each trap is characterized by $h_{trap}$, its vertical location in the
pixel and $\tau$, its characteristic release time constant. Different species of
charge traps have different decay time constants and thus different values of $\tau$ (and $\rho$).
Whenever a trap in a pixel containing $n_e$ electrons (or holes) is within the electron cloud, \ie\ $h_{trap}<h(n_e)$,
we assume that it immediately absorbs one electron from the free charge. When a full charge trap is above
the free charge height (that is, not within the electron cloud), it is given an opportunity to decay.

\begin{figure}[tb]
\plotone{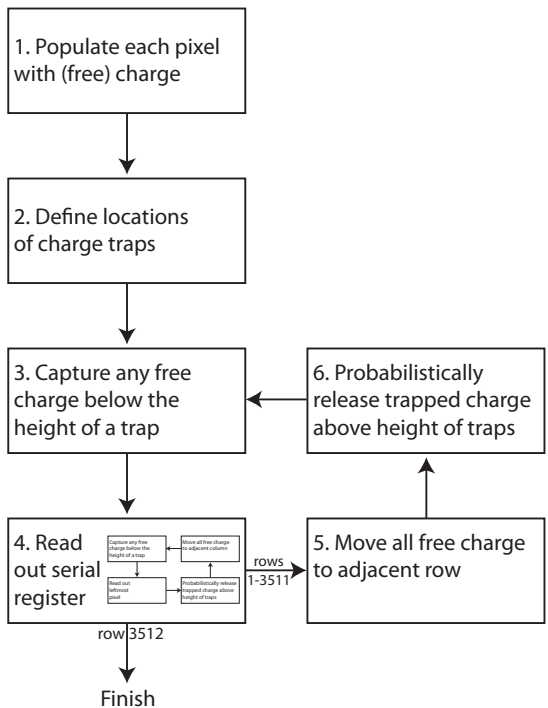}
\caption{Flowchart representing the  readout process as implemented in our code.
}
\label{fig:flowchart}
\end{figure}

We have developed a code that mimics the readout of a CCD with
imperfect CTI.  For each column of pixels, we use the following procedure
to ``read out'' the image and determine the observed charge in each pixel. This procedure is also illustrated graphically in  Figure~\ref{fig:flowchart}.
\begin{enumerate}
\item Populate each pixel with free charge, as accumulated during an exposure, and calculate $h(n_e)$.
\item Define the locations of charge traps throughout the pixel array. This is done for multiple  species of charge traps, each with a different $\rho$ and $\tau$.
\item For charge
 traps with $h_{trap}<h(n_e)$ fill the trap with one electron, and  subtract one electron from the free charge in its pixel.
\item Read out the charge in pixel row $n=1$.
\item For $n=1$ to $n=n_{max}$ set the free charge in pixel $n$ to the free  charge in pixel $n+1$.
\item For full traps with $h^{trap} > h(n_e)$, calculate the probability that the trap  will decay (that is, release the charge), based on an exponential  decay with time constant $\tau$.  Generate a random number in the range [0,1] and,  if the probability is less than the random number chosen, empty the trap and  increase the free charge by 1 electron.
\item Repeat the previous four steps to calculate the measured charge  in pixel rows $n=2$, $n=3$, ....  Note that $n_{max}$ is decreased  by one for each iteration.
\end{enumerate}

The image is then rotated by ninety degrees and exactly the same process is repeated to simulate serial transfers.
The total number of operations to read out one row or column of $n_{pix}$ pixels is $(n_{pix})(n_{pix}+1)/2$.
Thus, reading out out an entire $n_{pix} \times n_{pix}$ array scales as $n_{pix}^3$ and is computationally intensive.  During a readout, a charge can be trapped multiple times in different pixels.

This process inevitably adds noise to an image because we do not know the true locations of individual traps within the
detector\footnote[10]{This knowledge may be possible in future analyses using ``pocket pumping'', as described in
\S\ref{sec:future}.}, but simply model the traps as a uniform density. We therefore ought to perform many iterations and
average the results. As a more computationally efficient solution to this problem, we instead introduce ``fractional
traps''. That is, we place the same number of traps in each pixel and let each trap capture a ``fractional electron''
(with the fraction being $\rho$, the trap density per pixel, which can be less than $1$). We further divide the fractional trap in each pixel into $n_{levels}$
fractional traps, each located at a vertical position in multiples of $1/n_{levels}$. These traps release charge
exactly as described above. We have found that setting $n_{levels}=10000$ allows us to reproduce the averaged results of
many iterations with full traps placed randomly within the pixels (mimicking a real, physical CCD). However, the
fractional trap method saves considerable computational overhead when simulating the effects of CTI on many thousands of
 images as described in \S\ref{sec:simulations}.  Our code allows each trap to have a different $\tau$.
However, as discussed below, we find that in each charge transfer direction, a small number of $\tau$ values
(trap species) describe the behavior of physical detectors.

\section{Validation of the Charge Transfer Code}\label{sec:validation}

The charge transfer code has been tested and validated by creating simulated
images designed to mimic the irradiated data used in D08. Using the
same software as D08, we show that the code is able to reproduce the
observed radiation damage effects as measured in the D08 data.

\subsection{Irradiation in the Lab to simulate Space Radiation}



The damaging particle radiation flux incident on a spacecraft depends on exactly when the mission occurs within the
$\sim11$ year solar cycle (Barth \etal\ 2000).
There is an order of magnitude difference in the flux of solar protons during the heaviest and lightest parts of the
solar cycle (see, e.g. Figure~1 of Barth \etal\ 2000).
 In D08, the solar proton flux was modeled using the European Space Agency's Space Environment Information System (SPENVIS)\footnote[11]{See http://www.spenvis.oma.be}. In SPENVIS, a simplified  solar cycle consisting of 7 years near the maximum flux level and 4 years at zero flux is used. The total displacement damage (energy deposited in the silicon) is predicted from a uniform 4$\pi$ steradian spatial distribution of solar protons with an energy distribution derived from the Xapsos \etal\ (1999) model for solar proton emission. SPENVIS employs a statistical model  based on data from previous solar cycles to predict the dose at $95\%$ confidence level (CL); that is, the prediction will underestimate the dose only $5\%$ of the time. D08 used the SPENVIS model to calculate the 95\% CL solar proton flux incident on the CCDs, after passing through the shielding provided by the SNAP spacecraft and telescope, yielding  an integrated NIEL exposure
of $2.54 \times 10^6$ MeV/g (Si) for one year at solar maximum.


D08 then characterized the CTE performance of thick, fully depleted
LBNL p-channel CCDs by irradiating several CCDs at the LBNL 88-Inch
Cyclotron with $12.5$  and $55$ MeV protons. Although a variety
of irradiation levels were used, we only consider the $12.5$ MeV data with an irradiation level of
$2\times10^{10}$ protons/cm$^2$.  The NIEL factor for  $12.5$ MeV protons is $8.9\times10^{-3} \textrm{MeV}\textrm{g}^{-1}\textrm {cm}^{-2}$
per proton (Jun \etal\ 2003).
Thus, the $2\times10^{10}$ protons/cm$^2$ flux of 12.5 MeV protons used to irradiate the CCDs corresponds to $1.78 \times 10^{10}$ MeV/g (Si,) a total dose equivalent to ten solar cycles at 95\% CL, using the SPENVIS approximation detailed here, or 110 years at L2.

Although this accumulated radiation exposure is significantly
higher than any proposed dark energy mission would encounter,
the exaggerated radiation exposure makes it easier to characterize the detailed effects of CTE degradation.
Since the number of traps (and thus the degradation of CTE)
is linear with radiation exposure and the NIEL dose, the D08 accumulated radiation exposure can be used to estimate CCD
performance over the course of a dark energy mission lifetime.



\subsection{Analysis of CTI Due to Radiation Damage}

The CTE of irradiated CCDs was measured using a $^{55}Fe$ x-ray source
that emits K-alpha photons with energies of 5.9KeV. At the operating temperature
of 133K, a single K-alpha x-ray will generate 1580 electron/hole pairs,
which, depending on the location of the x-ray relative to the pixel
 potential wells, may be localized in a single pixel or shared among two
 or more pixels.


In D08, CTE was characterized using single pixel events
from the K-alpha peak, and the
results showed that the irradiated LBNL CCDs
are three times more affected by charge trailing in the parallel readout direction that in
the serial readout direction.
In this article, we disregard the serial CTE
and only consider the trailing in the parallel direction since this will
most affect galaxy shape measurements.

\begin{figure}[tb]
\plotone{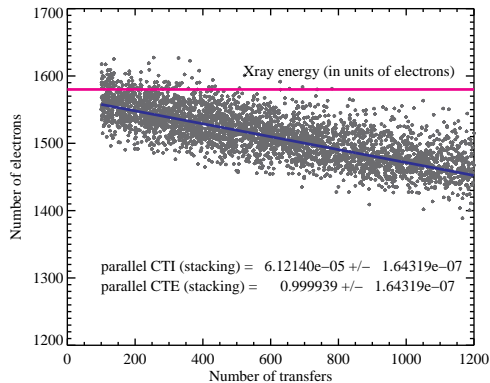}
\caption{Decrease in energy of single pixel events as a function of
  the number of transfers in the simulated images. The magenta
  solid line shows the mean energy of the simulated K-alpha impacts
  (1580 $e^-$). Because of charge trailing, the measured energy of
  these events after readout decreases with the number of
  transfers. The best fit line to the data is shown by the blue solid
  line and yields a CTE of $0.999939$.
}
\label{fig:test_cte2}
\end{figure}

\begin{figure*}[tb]
\plotone{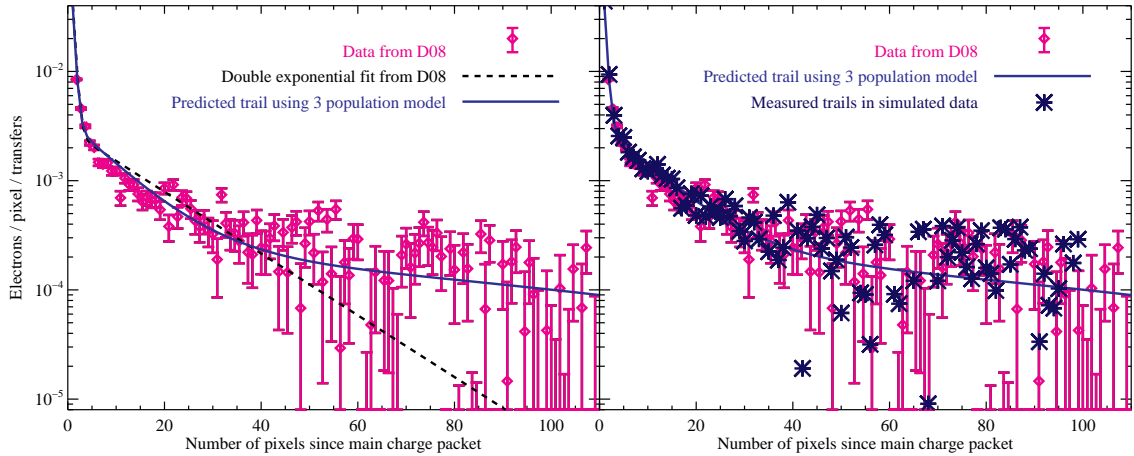}
\caption{This plot shows the  trailing pixel charge after 1650 transfers relative to the charge
  in the initial charge packet (the first pixel of an x-ray event). Left panel: Charge trailing measured behind single pixel
  events in the parallel read-out direction from D08 (combined data
  set from 14-17  August 2006) (magenta diamonds). The
  trails in this plot have been calculated up to 100 pixels after the
  main x-ray packet as opposed to 45 pixels in Figure 5 of D08. The
  dashed black line shows the two trap model used in D08 that
  underestimates the charge trapped in the far reaches of the tail. In
  order to best fit the data over the range 0 to 100 pixels, we use a
  3 population model and we determine the trap densities and time
  constants that best match the data. The predicted trail from our
  model is shown by the blue solid line. Right panel: We create a
  series of images designed to mimic the D08 data in terms of the
  background noise and the K-alpha impacts (signal). We read out the
  simulated data with imperfect CTE and measure the charge trailing
  using the same software as D08. The trails measured in the simulated
  data are shown by the dark blue asterisk data points and are a good
  match to the real data (magenta diamonds).
  The error bars for the simulated data would be about the same size as for the real data; for clarity the simulated error bars are not plotted.}
\label{fig:test_cte1}
\end{figure*}

The effects of
irradiation on CTE was studied  in two ways in D08.
In the first method, called the $\it stacking$ method,
CTE is characterized by the average charge collected for single pixel x-ray events
as a function of the number of pixel transfers.
Those x-ray events that
experience more transfers lose a larger amount of
charge due to CTI, as shown in Figure~\ref{fig:test_cte2}.  The serial and
parallel CTE components are determined independently by fitting the
fractional loss of each transfer to the data. The same single pixel
x-ray events  are used for a measurement of CTE using the
$\it trailing$ method, in which
the charge is counted in each trailing pixel as a fraction of the
charge in the primary charge packet.
The fractional trailing charge in each event is divided by the total number of transfers, and
the results are averaged
over all x-ray events. In other words, the averaged trails
represent the fraction of charge left behind the primary charge packet for a single transfer.  The effect of this trailing with a best fit to the data after irradiation and 1650 transfers is shown in Figure~\ref{fig:test_cte1}.   The total fractional
charge integrated over these trails represents the CTI.  To summarize
and compare the two methods, the \emph{stacking method offers a direct
measurement of  CTE}, measuring charge that is successfully transferred  relative to the expected x-ray
charge deposition,  while the \emph{trails method offers a direct measurement of CTI} by measuring the trailing charge relative to the charge in the leading pixel.
The total fractional
charge integrated over these trails represents the CTI.

In practice, the analysis of the trails following x-ray events is limited
by the ability to measure the faint trails after a large number of transfers
in the presence of non-zero read noise.
Because of the low S/N at large distances from the primary x-ray event,
D08 fit only the first 45 pixels with a two term exponential.
The charge in this two-term exponential represents approximately $2/3$ of the
total charge lost due to CTE effects as identified in the stacking method.  In a re-analysis
of the D08 data, in which stacking plots were made with
varying x-ray flux, we find strong evidence
that the remaining charge must be attributed to
one or more populations
of traps with a much longer time constant.
One such candidate is the C-O trap which has
been independently identified in the LBNL CCDs in a previous analysis (Bebek \etal\ 2002)
with a de-trapping time constant of many seconds compared to the typical
time between pixel row transfers of 25 ms at the $70,000$ pixel/sec readout speed employed in D08.

We therefore choose to model the CTI using three
distinct trap populations instead of the two used by D08.
We assign the third trap a time constant corresponding to 200 pixels in our analysis.
The results of our fit are shown in Fig. \ref{fig:test_cte1} and
the best-fit parameters are found to be: $\rho_1=0.35$,
$\rho_2=0.49$, $\rho_3=0.7$, $\tau_1=10.0$, $\tau_2=0.486$, $\tau_3=200.0$,
where $\tau$ is in units of `pixels'.  These best fit parameters are tabulated in Table~\ref{tab:params_tab}. There are, on average, about $1.5$ traps per pixel when summed over the three species; a mission that was at L2 for only about 5 years would thus have only about one twentieth that number of traps.
We created a set of simulated images with the same characteristics of the D08 images to test our models.
Overall, the agreement is good, as shown in Figure~\ref{fig:test_cte1}.

\begin{deluxetable*}{ccccc}
 \tabletypesize{\scriptsize} \tablecolumns{57} \tablecaption{Parameters of Charge Traps\label{tab:params_tab}}

\startdata
\hline
\hline
   Trap Species &      $\rho$ (traps per pixel) &  $\tau$ (pixels) & $\tau$ (ms)  & probable defect type  \\
\hline
1         & 0.35     & 10.0 & 250       & Carbon-interstial (Ci)    \\
2         & 0.49      & 0.486 & 12   &  divacancy (VV)   \\
3         & 0.7      & 200.0 & 5000   &   carbon-oxygen (CO)   \\
\enddata
\end{deluxetable*}

As we show in \S\ref{sec:simulations}, charge traps with
time constants that are long relative to the time between parallel transfers in readout
have negligible effect on galaxy shapes (i.e. they do not
leave trails). The trapped charge is, however, removed
from the object, and will thus affect photometry.  Because our primary
motivation is to study the impact of charge trailing on galaxy shapes,
we chose to mimic the measured trails in D08 instead of matching the
CTE as measured by the stacking method.  We note that the opposite
approach would be appropriate if our aim was to measure changes in
photometry instead of shapes.

\section{Application of the CTE Code to Simulated Galaxy images}\label{sec:simulations}

We create simulated galaxy images with de Vaucouleurs and exponential profiles and
use the code described in \S\ref{sec:code} to introduce the effects
of CTI on the galaxy images.  We
create galaxies as they would appear in the I-band of the proposed SNAP
mission, which has a 2 meter mirror, $0.1$'' pixels and
400 second exposures.  The background level is chosen to be the average
background for extragalactic observations taken from L2.  The measured background is slightly lower than the
input background because the CTI causes the flux to be dragged out into the overscan
region of the CCD during readout. We create single galaxies in each image to avoid having the traps in a pixel be filled by charge from an object that has already passed through that pixel during readout.
All objects are placed 1650 pixels from the readout register (close to the maximum of 1712 pixels from the readout register). This is done because objects farthest from the readout registers  will encounter the most traps and suffer the worst CTI; we are trying to estimate the worst case scenario for galaxy shape measurement.

%
%

We assume that the trap densities $\rho$ increase linearly over time, as traps
accumulate due to radiation damage, and that the time release constants $\tau$
do not change because they are properties of the detector material itself. The
assumption of linearity in trap density is not entirely accurate because the
proton flux is dominated by solar radiation, which varies over the solar cycle.
A specific analysis of any future dark energy mission will need to take into
account the portion of the solar cycle in which the mission occurs.    We make a further
simplification  by assuming that at the start of a mission a CCD
will have no charge traps, so only traps accumulated during the time spent at
L2 affect readout.  This is of course not true, because real CCDs always have
some imperfections even immediately after their production.  In the case of the
LBNL CCDs we are simulating, however, this turns out to be a good approximation because
the pre-irradiation CTE is so high and the number of traps so low (see Table IV
of D08). However, our results represent a best-case scenario for the number of
traps (and thus CTI) as a function of time; the real CTI will be slightly
worse.

For each galaxy, we measure the shape both before and
after the image is degraded with imperfect CTE.  The shape is
parameterized in the typical weak lensing fashion by a two component
ellipticity $e_i$, where $e_1=\frac{I_{xx}-I_{yy}}{I_{xx}+I_{yy}}$ corresponds to elongation along the $x$
axis (for positive $e_1$) or the $y$ axis (for negative $e_1$), and
$e_2=\frac{2I_{xy}}{I_{xx}+I_{yy}}$ corresponds to elongation at $\pm45$ degrees. Here, the shapes are described in terms of the second order moments of the pixel intensity $I$ such that $I_{ij}=\frac{\sum I w x_{i} x_{j}}{\sum Iw}$ where $x_i$ is the distance in pixels from the object centroid and $w$ is a Gaussian weight function.
The ellipticities
were measured using the method of Rhodes, Refregier, \& Groth (2000;
hereafter RRG). This method has been well-tested on real and simulated
space-based data (see  Leauthaud \etal\ 2007).  Since we are only
interested in perturbations to the galaxy shapes, we do not go through
the somewhat complicated steps of point spread function deconvolution,
which can introduce biases in shape measurements (Heymans \etal\ 2006;
Massey \etal\ 2007). Instead, we only concern ourselves with $\Delta
e$, which is relatively independent of the particularly shear
measurement method.  We explore the effects of S/N and galaxy size on
$\Delta e$ but for the bulk of our analysis, we consider small, faint galaxies;  any weak lensing survey
will be dominated by galaxies that are faint and small relative to the PSF size.

For the purposes of this paper, we only introduce
parallel CTI into the simulated images and we set the
serial CTE equal to 100\%.  We do this for two reasons.  First, the
parallel CTI smears objects in the vertical direction
(negative $e_1$), but serial CTI smears them horizontally
(positive $e_1$, the serial readout direction).  Thus, just using the
ellipticity $e$ as an indicator of CTI-induced galaxy shape changes
means that the effects of serial and parallel CTI partially cancel.
The combined effects change the \emph{size} of the PSF, and thus
must be corrected for in real images, but would provide an unfair test for these purposes.  The second reason we
concentrate just on parallel CTI is that the parallel CTE
degradation is three times worse for a given radiation exposure (D08) and
different charge trap species affect the parallel and serial CTE
because of the different clocking times in the parallel and serial
directions.  Thus, we seek only to demonstrate that we understand the
more influential effects of the parallel CTI on shape measurement in this paper.

\subsection{Effect of charge trap release time}\label{sec:tau}

\begin{figure}[tb]
\epsscale{0.7}

 \plotone{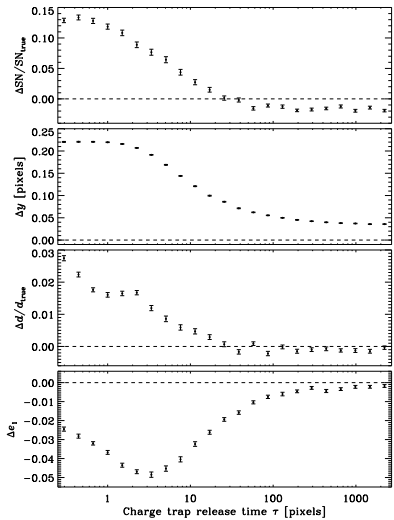}
  \caption{The effect of charge trap release time $\tau$ on a
    measurement of photometry, astrometry, size and ellipticity of a
    typically small, faint galaxy degraded by CTI. Each $y$ axis represents the fractional change in that quantity. The absolute values
    of the $y$ axes are largely irrelevant, depending upon the assumed
    density of charge traps, CCD well filling model, galaxy SN (13),
    size (FWHM=3.8 pixels) and morphology (circularly symmetric De
    Vaucouleurs profile). However, the trends reveal several lessons
    for future hardware: notice particularly the local maximum in
    $|\Delta e_1|$, which implies a worst case clocking time, or readout
    cadence, for CCDs.}
\label{fig:tau_effect}
\epsscale{1.0}
\end{figure}

Figure~\ref{fig:tau_effect} demonstrates the effect of charge trap release time
$\tau$ on the measurement of photometry as measured by S/N (calculated via
SExtractor; Bertin \& Arnouts 1996, S/N =flux\_auto/fluxerr\_auto), astrometry (the $y$ centroid of an
object), and shapes (rms size $d_{rms}=\sqrt{0.5(I_{xx}+I_{yy})}$ and ellipticity
$e_{1}$). We measure the release time $\tau$ in
units of ``pixels,'' the amount of time it takes to clock the charge by a certain
number of pixels in the parallel readout direction (i.e., one ``pixel'' is the
time between row shifts during readout).


In terms of astrometry, photometry, and size, there are two limiting
regimes. Charge traps with very short release times (or slow CCD
readout) push charge from an object's leading edge onto its core,
and drag core charge into a short tail. Both effects shift the
object away from the readout register. The net effect also increases
the object's size, because the core contains more charge than the
wings. A small amount of flux can be lost from the wings into a tail,
so $\Delta$flux is always slightly negative. However, the smoothing
inherent in trailing correlates adjacent pixels and has the perverse
effect of {\em increasing} the  S/N. Note that the limiting
behavior at low $\tau$ is as expected: in our model, all traps inside
a charge cloud carry an electron to the adjacent pixel at every clock
cycle. In a real CCD, some charge may be released from very fast
charge traps part-way through the 3-stage clocking cycle and returned
to their original pixel. This process would lower the effective
density of charge traps with low $\tau$.

Charge traps with long release times (or fast CCD readout) steal flux
primarily from an object's leading edge, and return it to the image in
pixels well separated from the object. This stolen flux lowers the
detection S/N. It also shifts the centroid as before, and
decreases the size. For intermediate $\tau$, these effects are
dominated by the addition of a tail, which increases the overall size.
One curious dependency upon measurement method is that, while the rms
size $d_{rms}$ decreases with $\tau$,
the FWHM fitted by {\tt SExtractor} {\em increases}: for example, $
\Delta$FWHM is negative for small $\tau$. This is presumably related
to the net increase in detection S/N, and the segmentation
of the image into fewer pixels that {\tt SExtractor} determines belong
to a given object. Note that if high $\tau$ were achieved by
dramatically speeding the CCD readout, our assumption of
instantaneous capture times may become invalid. A probabilistic
capture mechanism over a finite time would result in lower effective
densities of all traps, and potentially increased sensitivity to the
density of charge throughout a pixel potential, changing the well
filling parameters $\alpha$ and $d$.

The spurious ellipticity induced in an object is interestingly
different. The tail and the centroid shift induced by charge traps
with short release times both  elongate an object in the
readout direction. As the tail lengthens, the spurious ellipticity
initially
increases. However, once the charge in the tail is sufficiently disconnected from
the object and the
object's centroid shifts back towards the correct position, the
spurious ellipticity begins to decrease. In the limiting case of
charge traps
with very long release times, charge missing from the object's
leading edge could potentially elongate the object perpendicular to
the readout direction; however, the residual centroid shift in this
case is sufficient to maintain a small ellipticity in the readout
direction (this result may depend upon the object's radial profile).

Thus, we find that, in terms of weak lensing shear measurement, not all CTI is
equally bad. Furthermore, there is a worst possible case, in which traps with
release times corresponding to 3--4~clock cycles induce the most spurious
ellipticity. This value depends upon the shape measurement method: with KSB (Kaiser, Squires \& Broadhurst 1995) and
RRG, it depends upon the size of the Gaussian weight function. The bump in
$\Delta d(\tau)$ around this value is real and also depends upon this scale.
However, from a more general argument about the dissociation of flux from an
object in a very extended trail, it is clear that a local maximum in $|\Delta
e_1|$ will be inevitable for all shear measurement methods. The
clock speed is a parameter that can be tuned in the hardware.  We discuss this
possibility in \S\ref{sec:future}.

\subsection{Effects on galaxy morphology}

\begin{figure*}[tb]
 \plottwo{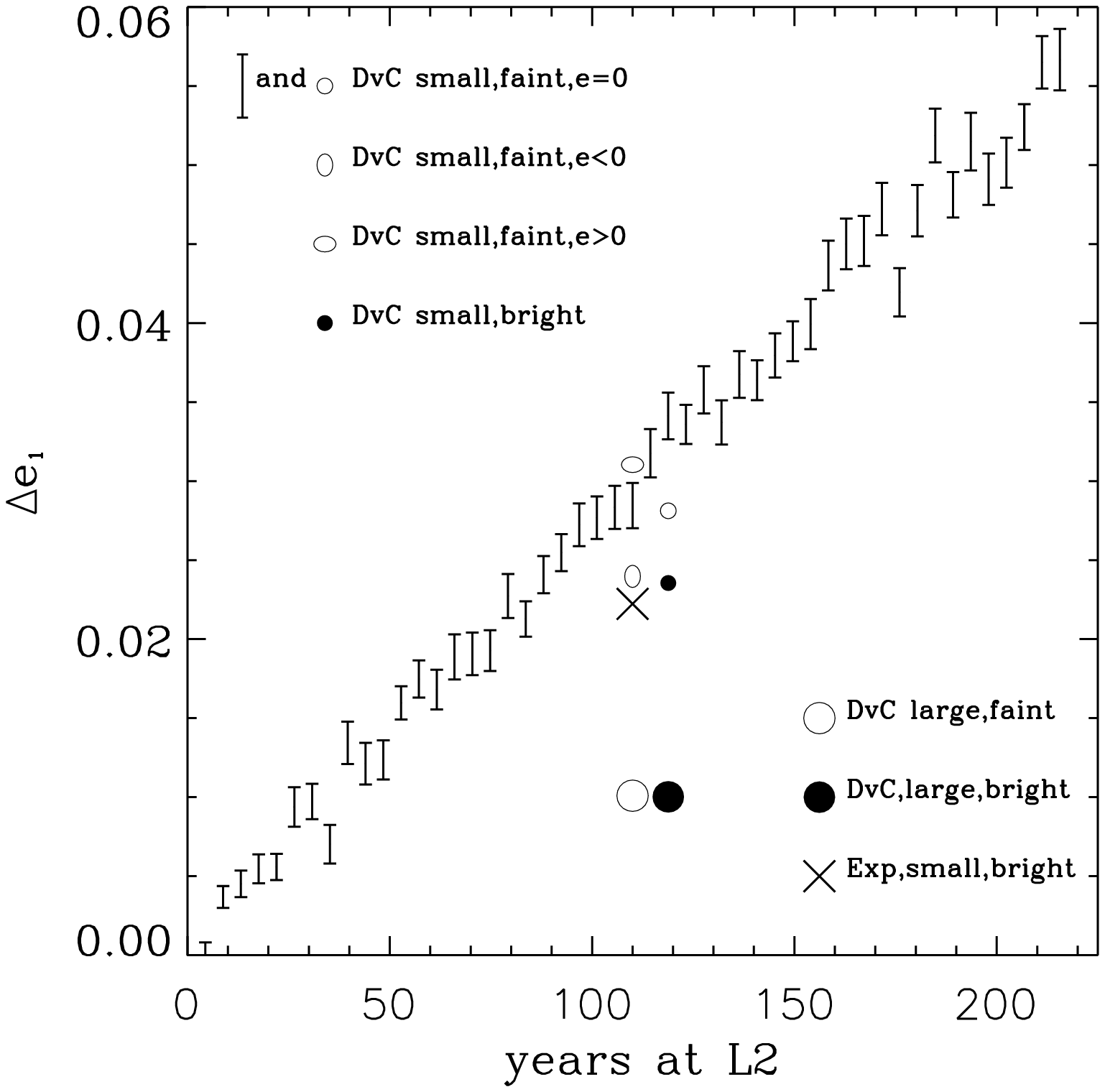}{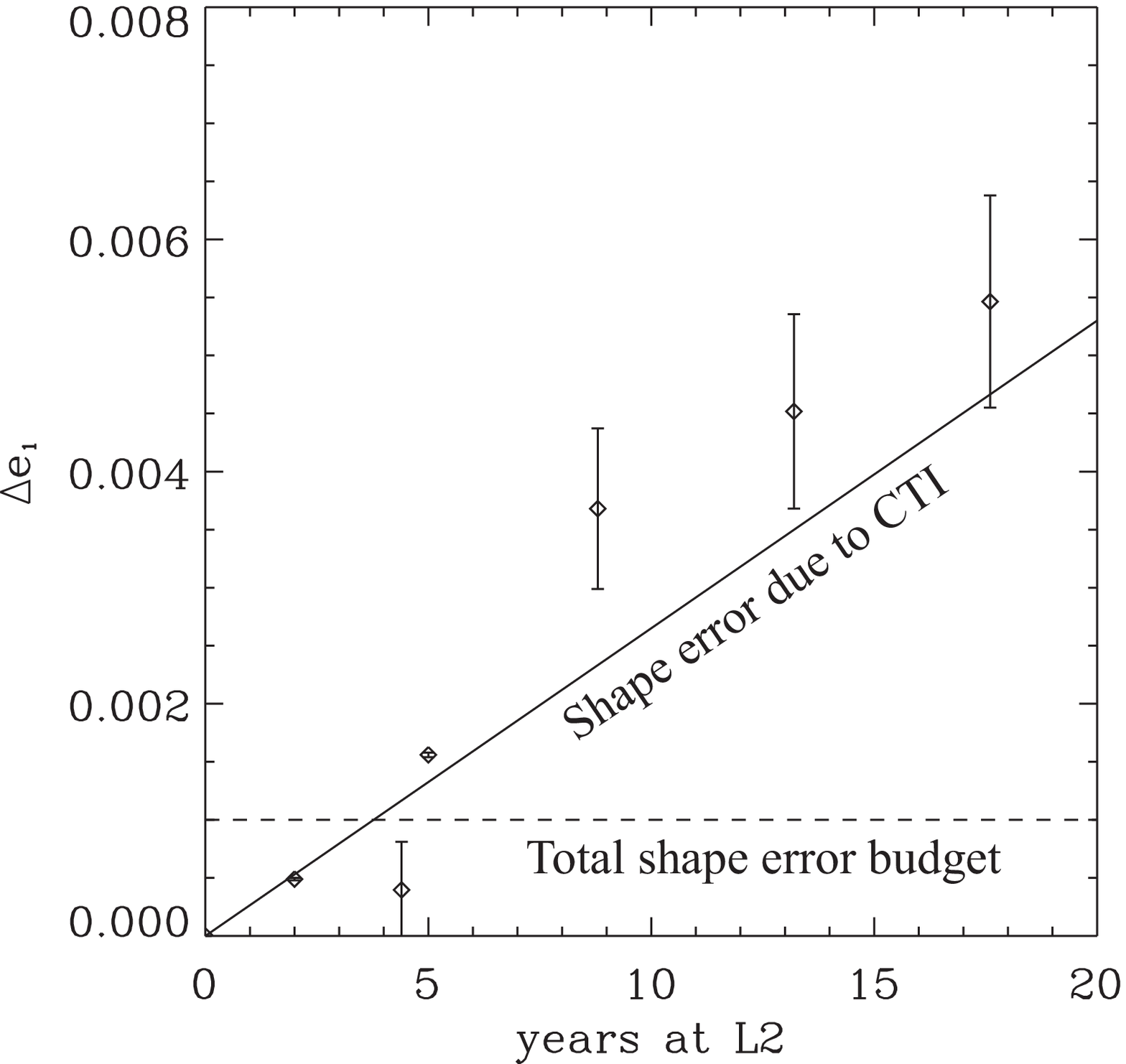}
  \caption{\textbf{Left Panel}: The degradation in shape measurements ($|\Delta e_{1}|$) as a function of time at L2.  We make simulations for radiation exposure from 0 to 220 years at L2 (the latter number being twice the accumulated radiation exposure used in D08).
  The data points with error bars are small, faint De Vaucouleurs
  profiles far from the readout register, and therefore represent a worst-case scenario for CTE-induced shape errors. The other data points represent different levels of SN, size, and profile type as described in Table~\ref{tab:simstable}. We create and read out 3000 galaxies of each type, so that the error bars on those points are smaller than the plotting symbols.  All of the individual profile types are plotted for 110 years of L2 radiation exposure; some points are slightly offset for readability. \textbf{Right Panel}: An enlarged portion of the left panel showing the effects of radiation damage in the first 20 years at L2. The data points at 2 and 5 years are described in \S\ref{sec:future} and have error bars about the size of the plotting symbols.  The solid line is our best fit to our data points as given in Equation~2. The dashed horizontal line is at $\Delta e=0.001$, the level of shape measurement accuracy needed for a future all-sky dark energy mission (Amara \& Refregier 2008).}
\label{fig:trap_density}
\end{figure*}

We create a series of images at the fiducial irradiation level of D08 (110
years at L2).  We created galaxies with De Vaucouleurs (DvC) profiles and
exponential profiles. For the DvC galaxies, we varied S/N, size, and input
ellipticity. For each different set of  simulation parameters, we create
1000 simulated galaxies, each with a different, random sub-pixel position of
the galaxy centroid and different background noise realization. We measure the size in terms of the rms
size $d$. Small
galaxies have a size close to that of the PSF, representing the typical
galaxies that will dominate a lensing survey; large galaxies are significantly
bigger than the PSF. The values of S/N, size, and ellipticity for the different
simulations  are shown in Table~\ref{tab:simstable}.
The key result for weak lensing, the change in
measured ellipticity, is illustrated in Figure~\ref{fig:trap_density}.


\begin{deluxetable*}{llccccc}
 \tabletypesize{\scriptsize} \tablecolumns{57} \tablecaption{Parameters of Simulated Galaxies\label{tab:simstable}}

\startdata
\hline
\hline
   Profile &      S/N &  $\Delta$(S/N) & $e$  & $\Delta e$ & rms size $d_{rms}$ [pixels] & $|\Delta y |$ \\
\hline
DvC         & low (13)      & 1.1 & 0       &   -0.028    &  (small) 1.8   &   0.19  \\
DvC         & low (13)      & 1.1 & -0.17   &    -0.024   &  (small) 1.8   &   0.19  \\
DvC         & low (13)      & 1.2 & +0.17   &  -0.031     &  (small) 1.8   &   0.19  \\
DvC         & high (50)     & 0.7 & 0       &   -0.024    &  (small) 1.9   &   0.17  \\
DvC         & low (20)      & 0.3 & 0       &   -0.010     & (large) 3.1    &  0.20   \\
DvC         & high (50)     & 3.0 & 0       &   -0.010     & (large) 3.6    &  0.20   \\
Exponential & high (49)     & 0.9 & 0       &   -0.022    &  (small) 1.9   &   0.18  \\
\enddata
\end{deluxetable*}

As expected, the small galaxies are significantly more
affected by CTI than large galaxies. We also show that for small galaxies,
brightness (S/N) is a mitigating factor (but not for large galaxies); small bright objects are slightly less
affected by CTI than small faint ones.  Another interesting feature recovered from
these simulations is the dependence of $|\Delta e|=|\Delta e_{1}|$ on galaxy ellipticity.
Galaxies that are already aligned along the readout direction ($e_{1}<0$ in the
case of our simulations) are less affected by CTI than galaxies that are
aligned perpendicular to the readout direction ($e_{1}>0$).  Perturbations in the
$y$ direction (such as CTI) affect galaxies that are already aligned in the $y$
direction less than galaxies aligned in the $x$ direction.

\subsection{Effects of trap density}\label{sec:plots}

Figure~\ref{fig:trap_density} also shows the results of simulations with
varying trap densities.  We increment the trap density (in all three species)
from zero to the density that would be found after 220 years at L2 (twice the
fiducial value from D08).  At each of 50 evenly spaced points along this
timeline, we create 100 simulated galaxies, each with a DvC profile (small,
low S/N, $e=0$).  As expected, the degradation of shape measurement increases
linearly with trap density.  We find that
\begin{equation}\label{eqn:delta_e_rate}
\frac{{\mathrm d}\Delta e}{{\mathrm d}t}=(2.65\pm 0.02)\times10^{-4} \textrm{~[yr}^{-1}\textrm{~at L2]}
\end{equation}
\noindent for a radiation dosage averaged over an 11-year solar cycle and yearly displacement damage dose of $1.6 × 10^6$ MeV per gram of
Si.
We make the simplifying assumption of zero traps (CTI$=0$) at time $t=0$.  However, D08 show that
the number of traps in a new LBNL CCD is smaller than the measurement error, so the approximation we make here is a good one.

\section{Consequences for the Future Space Missions}\label{sec:future}

Amara \& Refregier (2008) ascertain that the multiplicative error on measured
shear needs to be kept below one part in $10^{-3}$ in order that future dark energy missions
not be dominated by systematic errors.  This means that \emph{all}
sources of shear measurements error (not just the portion due to CTI) must be
kept below $\Delta e<10^{-3}$ throughout the mission lifetime. This level of shape measurement accuracy is
represented in Figure~\ref{fig:trap_density} by a horizontal dotted line.

The prediction for $\Delta e$ after a fiducial 5~year mission can be calculated from
Equation~(\ref{eqn:delta_e_rate}). To avoid any assumption of linearity with $\rho$, however,
we have also run a larger number of simulations at 2 and 5 years of mean L2 exposure.
In each case, we created 3000 simulations to reduce measurement noise due to the sub pixel
galaxy position and sky noise.
We find that $\Delta e_{\textrm{2
years}}=0.490\pm0.01\times10^{-3}$ and $\Delta e_{\textrm{5 years}}=1.56\pm
0.02\times 10^{-3}$. That is, without any correction,
a 5 year weak lensing mission will have the entire
shape measurement error budget consumed by CTI-induced effects before the end
of the mission, even with specially designed, fully depleted, radiation
hardened, p-channel CCDs.

Fortunately, recent work using data from the HST's Advanced Camera for Surveys Wide
Field Camera (ACS/WFC) has shown that software postprocessing can correct the effects of CTI on galaxy
shapes by about a factor of 10 (Massey \etal\ 2009). Using the same code described here, trailed charge
could be moved back to where it belonged in an iterative procedure  restoring images
to their true appearance. Massey \etal\ (2009) measured the time constants and number density of traps
in ACS/WFC as a function of time, using extragalactic survey imaging that would be naturally available
in any future survey without additional overhead. For this software, the factor of ten level of correction will be
maintained down to the regime of future missions with much lower trap densities. One component of noise
(due to variations in the number of traps in a given pixel, which we treat as a constant density) will be
improved, but this affects only scatter in $\Delta e$ rather than the level itself. We therefore
conclude that, \emph{using CCDs with the characteristics of those in our study, and proven software mitigation
techniques to achieve an additional factor of 10 correction, CTI in a future dark energy mission would
be satisfactorily controlled at only 10\% of the total shape error budget.}
\\

One way in which CTI models (and mitigation techniques) could potentially be pushed beyond correction by
a factor of 10 would be to precisely locate individual charge traps, rather than treating them
statistically.  This would be most beneficial in the early years of a dark energy mission, when
$\rho << 1$. Designing flexibility into the clocking speed, waveform, and voltage
in CCD electronics provides the ability to locate traps via ``pocket
pumping'' (Janesick 2001).  Pocket pumping is a process in which a
uniform level of charge introduced by a flat field lamp is rapidly
shuffled back and forth thousands of
times in the parallel direction. Following the charge shuffle, the
charge is transferred to the readout transistor in the normal manner.
The resulting image reveals accumulated charge captured and then
released by each trap in the shape of a dipole of an overdensity
neighboring an underdensity of charge.  The orientation and strength
of this dipole reveal the location of the trap within the pixel and
the effectiveness of the trap. This can be repeated with different
levels of initial charge to map out traps in 3 dimensions within the
CCD.  Because the readout time of the CCD is dominated by the clocking
of serial charge, the pumping of charge in the parallel direction does
not introduce a significant amount of overhead to the survey.  For
example, it takes approximately 45 seconds to record a normal image
from the LBNL CCDs using the D08 clocking parameters.  Image
acquisition takes an additional 45 seconds for 20000 cycles of pocket
pumping with a five pixel shift.  Acquisition of five successive
pocket-pumping images every week to reject cosmic rays and model the
traps therefore would account  for less than ten minutes of additional
calibration time.

Another suggested mitigation technique is ``charge injection'' (also called ``fat zero'' or ``pre-flash'') in which charge is placed into the pixels in order to fill the traps.
The charge injection is simply a method to
increase the overall background level and fill the volume.  However, this has the effect of increasing the background in an image (much like increasing the zodiacal background) and will reduce the S/N of the detected objects.  This is obviously undesirable for a weak lensing experiment in which the observer is attempting to measure shapes of faint galaxies.

We showed in \S\ref{sec:tau} that there exist clocking time scales that are maximally bad for shape
measurement.  If the clock cycle is 3--4 times the trap release time, then $\Delta e$ is maximized.
Thus, future missions with weak lensing as a primary science driver should include an optimization of
the charge clocking time in their CCD readout electronics.
Increasing the rate at which charge is clocked serially, and thus increasing the rate at which parallel transfers of rows can be made, increases readout noise, resulting in an effective loss of survey depth; decreasing the rate at
which charge is serially clocked increases the readout time and, if that dominates over factors like slew and
settle time between exposures, will reduce survey area.  Thus, careful consideration must be paid to the
trade-offs in any such optimization.  Furthermore, we are only making recommendations for how to
minimize shape measurement errors due to CTI. Photometry is also degraded by traps with large values of
$\tau$ relative to the charge clock period. This can remove charge from objects but place it far enough away
that the shape is not significantly affected. Thus, trade-offs in charge clocking time must also take
into account the photometric accuracy requirements of the mission.


We finally note that the temperature at which detectors are operated at has significant effect on CTE, and thus future missions should be tested and optimized with this in mind.


\section{Conclusions}\label{sec:conclusions}

We have quantified the effect of CTI on measurements of weak gravitational lensing.
We first simulated the transfer of charge within LBNL p-channel CCDs using a model for charge traps
with three characteristic release times to reproduce the experimental results of D08.
Using this model, we then simulated deep exposures of galaxies for weak lensing measurements
from a space-based telescope subject to radiation damage.
The resulting simulated data were used to quantify the
effects of radiation damage on shape measurements of galaxies of various sizes and S/N levels, the true
shapes of which were precisely known. Most galaxies in any weak lensing survey will be small and faint;
as expected, we find that these suffer worst from the effects of CTI.

The level of CTI-induced shape error $\Delta e$ will approach the total shape error budget of a dark
energy mission (1 part in $10^{-3}$; Amara \& Refregier 2008) after less than 4 years of radiation exposure at L2.
Software mitigation techniques in image postprocessing, already proven on HST data (Massey \etal\ 2009),
will be able to reduce the levels of shape error well below mission requirements. We have also suggested
hardware capabilities, such as ``pocket pumping'' and adjustments to the readout speed, that may provide
additional help. However, our numerical results are only valid for p-channel devices, whose CTE characteristics after
radiation exposure have been shown to be superior to more common n-channel devices (Lumb 2009; Marshall \etal\ 2004).
\emph{Given the necessity of both hardware and software mitigation of CTI effects for successful mission
operation, we recommend that future spacecraft be designed with detectors and mission parameters that ensure CTE characteristics
no worse than the LBNL p-channel devices we simulated for this work.}\\

There are several caveats to our results.  First, we assumed that all galaxies are small, faint, and lie
far from the readout register.  Galaxies that are bright, large, or nearer  the readout register will
suffer less from CTI. Indeed, the average distance to the readout register will be exactly half of the
worst case scenario we outlined, so the mean $\Delta e$ will be a factor of two lower. In a real mission, there will be several dithered exposures of each galaxy, with each dither placing the galaxy a different distance from the readout register; this may allow us to further model the effects of CTI on shapes and partially mitigate those effects. We also assumed
a high level of radiation from the Sun for at least a portion of the mission due to modeling radiation flux in the heaviest part of a typical
11-year solar cycle.  A mission flown during the level of minimum particle radiation at L2 would suffer less radiation
damage, but a mission flown entirely during the maximum of the solar radiation flux would suffer more damage.
This is a large uncertainty because the radiation flux due to the Sun can vary by an order of
magnitude over the solar cycle (Barth \etal\ 2000). Furthermore, any real mission would need to adjust the flux according to the planned shielding
on the spacecraft (D08 assume the SNAP design) and should take into account secondary particle cascades
from reactions of high energy radiation with the shielding material (D08 ignored such secondary
radiation).
The `stacking' and `trails' methods probe different properties of CTE and we have chosen the fits to
trails in D08 to model the effects of radiation exposure on shape measurements.
We have only explored the charge transfer and radiation tolerance properties of a certain model of CCD operating at a single temperature;  any future space-based weak lensing missions should undertake a similar analysis using the CCDs planned for that mission.
Finally, if the CCDs contain a significant density of traps even before they are launched
into the harsh radiation environment of space, the CTI will be worse throughout, and the useful mission
lifetime reduced. Clearly, then, this paper is a first step, and any future mission should use a
procedure similar to the one we have developed in this paper to meet specific mission requirements by
fully optimizing its choice of CCDs, clocking rate, and shielding.

\acknowledgments

\noindent {\bf Acknowledgments}

This work was supported in part by the Jet Propulsion Laboratory,
operated by the California Institute of Technology under a contract
with NASA. This work was also supported
by the United States Department of Energy under contract
No. DE-AC02-05CH11231.
 We thank Chris Bebek, Mike Lampton, Michael Levi, and Roger Smith for useful discussions
about CCDs and CTE. AL acknowledges support from the Chamberlain
Fellowship at LBNL and from the Berkeley Center for Cosmological
Physics.
RM is supported by STFC Advanced Fellowship \#PP/E006450/1 and FP7 grant MIRG-CT-208994. CS was supported by funding from the Office of Science at LBNL and Fermilab.

 \end{document}